# Kinetically determined shapes of grain boundaries in CVD graphene

Ksenia V. Bets, Vasilii I. Artyukhov, Boris I. Yakobson*

Department of Material Science and NanoEngineering, Rice University, Houston, TX 77005, USA
*Corresponding author: biy@rice.edu

Predicting the shape of grain boundaries is essential to control results of the growth of large graphene crystals. A global energy minimum search predicting the most stable final structure contradicts experimental observations. Here we present Monte Carlo simulation of kinetic formation of grain boundaries (GB) in graphene during collision of two growing graphene flakes. Analysis of the resulting GBs for the full range of misorientation angles α allowed us to identify a hidden (from post facto analysis such as microscopy) degree of freedom – the edge misorientation angle β. Edge misorientation characterizes initial structure rather than final structure and therefore provides more information about growth conditions. Use of β enabled us to explain disagreements between the experimental observations and theoretical work. Finally, we report an analysis of an interesting special case of zero-tilt GBs for which structure is determined by two variables describing the relative shift of initial islands. We thereby present analysis of the full range of tilt GB ($\beta \neq 0$) and translational GB ($\beta = 0$). Based on our findings we propose strategies of controlling the GB morphology in experiments, which paves the way to a better control over graphene structure and properties for advanced applications.

## Introduction

The unique physical properties of graphene [1,2] invite many applications [3,4], but the only scalable method of graphene production—CVD—is always accompanied by grain boundary (GB) defects [5,6]. These defects affect the electronic properties [7,8], strength [9-11], and chemical stability [12]. For a long time presence of grain boundaries in graphene was treated as setback, but lately as GB's effect was understood to be very complex [13,14] researchers turned to seek possible advantages of having localized lines of defects in graphene. It was shown that specific structure of GB should be taken into account as effect would vary with change of defects arrangement [15,16] and selective use of GB may be used for specific purposes [17-20]. Knowing how grain boundaries form and what affects their shape is therefore instrumental in optimizing the quality of material.

Unfortunately, so far there is no theoretical base that would allow controlling shape of GB. Theoretical studies usually concern themselves with simple straight and periodic GBs [21,22]. Experimentally, meandering GBs with an unexpectedly high concentration of defects are often observed [23,24]. This is understandable as the final shape of GB is not dictated by the minimization of the *global* energy of the system, but rather, by the "near-sighted" *local* energetics of attachment of each subsequent atom [25]. As a consequence, a complete understanding of grain boundaries in graphene is impossible based on idealized/simplified structures, and calls for studies of the processes of their formation during graphene growth. In this paper we report on atomistic Monte Carlo simulations of GB formation in the course of an impingement of two growing graphene islands. After briefly introducing the computational scheme we present typical morphologies observed during simulations, with clearly different patterns of distribution and orientation of dislocations. These patterns are determined by a "hidden variable": the aperture angle between the edges of merging grains, which is in a many-to-one relation to the lattice misorientation angle and thus is non-obvious from post-growth observations of GB morphology. Finally we present analysis and the full range of structures of translational GB (zero edge misorientation) which are determined by two variables describing the relative offset of the islands.

## Monte Carlo growth simulation

Graphene growth was simulated using a Monte Carlo procedure. At each step, new atom additions were attempted at all edge sites (armchair, zigzag, and kink) as shown in **Fig. 1(a,b,c)** in the system. Rather than operating on a fixed 2D lattice, multiple possible additional configurations were sampled making sure that configurations containing defects (pentagons, heptagons, etc.) could form. To balance out accuracy and computational speed, the energies were calculated using the AIREBO interatomic potential [26]. The structure was constrained to two dimensions to mimic the presence of a weak substrate such as Cu. The carbon feedstock was assumed to be in the form of single atoms.

**Fig. 1d** shows the shape evolution for a single flake starting from an arbitrary configuration containing all types of edge sites. Armchair and kink sites present favorable bay regions where insertion of new atoms is easy, and as a result, the shape evolves into a kinetically-determined form of a hexagon with zigzag edges, as predicted theoretically [27] and observed in experiments [28] (**Fig.1d**). To further estimate accuracy of growth representation by our simulations we investigate process of filling of a hexagonal void in perfect graphene sheet (Fig. 1e). The edges of the void evolve to a ~19 degrees orientation relative to the zigzag direction, corresponding to the fastest-growing direction in graphene, as recently observed experimentally by Ma *et al.* [29]. Judging from similarity of our results to experimental data, our procedure correctly represents the growth of graphene under low supersaturation conditions

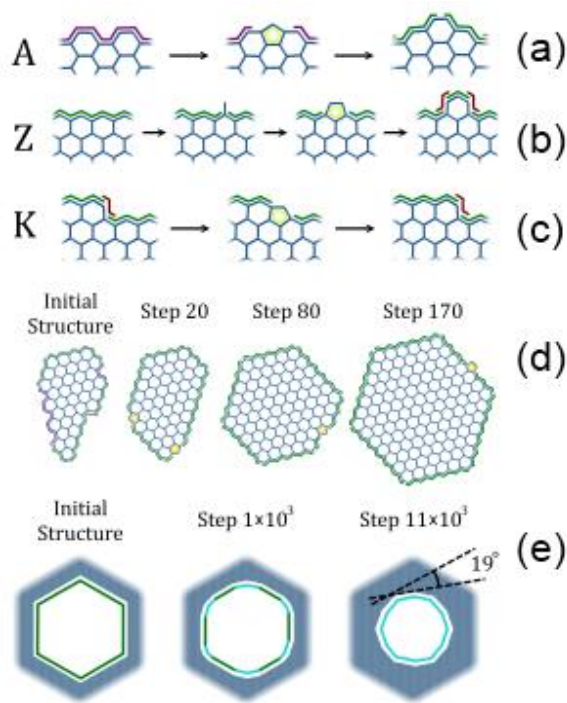

**Figure 1. Monte Carlo simulation algorithm**. Structures resulting of adsorption to different edge sites: **(a)** armchair, **(b)** zigzag, **(c)** kink edge. **(d)** Evolution of an arbitrary starting flake shape into the kinetically driven shape of a Z-edged hexagon. Similar process was shown in [27]. **(e)** Evolution of zigzag edges inside hexagonal void in graphene into edges with $19^{\circ}$ slope. Side of the hexagonal void in simulation is 1.7 nm. Similar process was shown in [29].

(before diffusion control sets in causing the formation of dendrites).

## Computational experiment

To observe the formation of a GB, we perform simulations of growth starting from two closely located hexagonal grains (represented by zigzag edges - **Figure 2a**) that grow concurrently until their edges meet and the islands start to merge producing a GB. The outside region (not participating in growth process) was fixed in order to prevent shift and rotation of the initial structure.

Edge misorientation β (**Fig. 2a**), an angle between merging zigzag edges in the initial structure, was used as main parameter. Compared to the lattice misorientation angle α (commonly used in literature), β provides additional information about the initial structure and therefore conditions of GB formation. Edge misorientation has a many-to-one relationship with lattice tilt (shown on **Fig.2b**) as α represents relative positioning of lattices of grains, while β determines the exact orientation of merging edges.

**Morphology analysis**

By varying edge misorientation (initial structure) we observed four distinct families of grain boundaries displayed in **Figure 3 (a-d)**. Observed types of GBs can be classified by its shape as: straight (**Fig.3a**) – with all atoms forming straight line along bisector of initial angle between edges of grains (from now on would be called bisector direction); scattered straight (**Fig.3b**) – usually formed by sparsely located defect pairs aligned with bisector direction, those defect couple can be slightly shifted from bisector direction; alternating (**Fig.3c**) – with defect pairs orientation switching between bisector direction (all 5-7 pairs in presented structure) and zigzag direction of one the grains closest to the bisector direction (all 7-5 pairs in presented structure); and curved (**Fig.3d**) – with defects following bisector direction, first zigzag direction of grains and second zigzag direction of grains. Such a striking difference in morphology needs to be explained, however first we need to establish a way of describing the morphology quantitatively. To approach that problem we consider orientation of pairs of closest defects (5-7, 7-5) that form GB as a chain; this way position of each next defect in GB would be considered in comparison to the defect that goes before it in the "chain". Using that logic it is easy to notice that each structure is different from previous one by additional orientation of defect pairs (highlighted on respective structures) and that orientation is responsible for over-all shape of GB.

As a measure of defect pair orientation we use its director angle γ, which is the angle between the line passing through the centers of the two consecutive defects, and the bisector of the starting aperture angle β, as shown in **Fig. 3e**. By computing γ for all sequential defect pairs we obtain a γ distribution for each structure, as shown in **Fig. 3 (f–i)**. Each peak on γ distribution plots correspond to one type of defect couple orientation and peaks corresponding to orientations highlighted on the structures (**Fig.3a-d**) are also highlighted on the plots (**Fig.3f-i**). Those principal orientations would be referred to as a center peak (**Fig.3a, f**), shift (**Fig.3b, g**), first rotation (**Fig.3c, h**) and second rotation (**Fig.3d, i**).

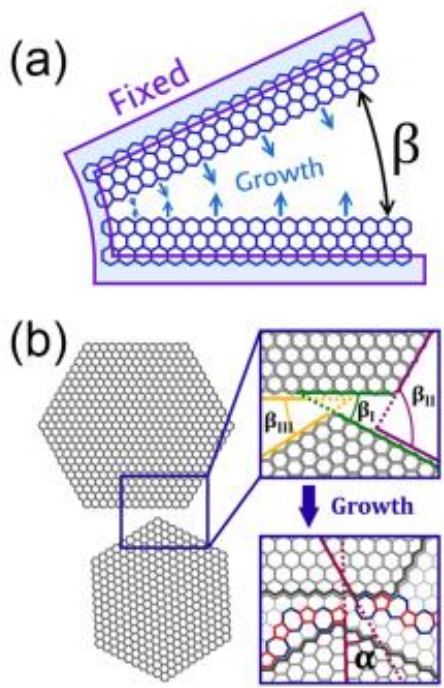

**Figure 2. Computational experiment of GB formation.** **(a)** Scheme of computational experiment showing edge misorientation angle β. **(b)** Example of relation between edge misorientation angle β and lattice misorientation angle α. Several values of edge misorientation angle β correspond to a single value of lattice misorientation angle α.

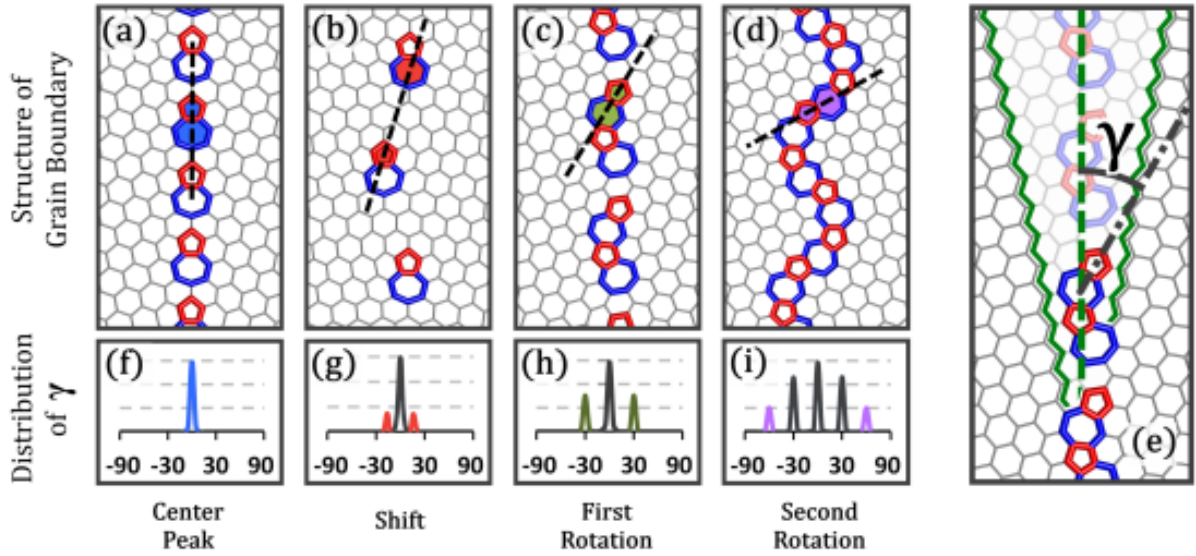

**Figure 3. Results of GB formation simulations**. Examples of a straight (**a**), scattered straight (**b**), alternating (**c**) and curved (**d**) GBs. Structures correspond to β = 81.2° (α=21.8°), 73.17°(13.17°), 21.8°(21.8°) and 86.8°(26.8°) respectively. Typical orientation of pairs of defects responsible for the formation of the corresponding structures is highlighted and direction of the pair (line going through centers of both defects) is shown. (**e**) Director angle γ of defect couples with respect to the bisector of β used to characterize orientation of each defect pair. β represented by highlighted zigzag direction and its bisector by green dotted line. (**f–i**) Distribution of γ (bisector direction is vertical) corresponding to the structures on (a-d). Peaks corresponding to orientations highlighted on respective structures are also highlighted on distribution plots.

Using structural content of GB – proportions of types of defect pair orientations in it, we can characterize the overall shapes of observed GBs. **Fig.4a** shows dependence of structural content of GB from edge misorientation angle β and misorientation angle α. Samples of actual structures are shown on **Fig. S2**.

As another characteristic of GB shape we calculated average direction angle for each value of β. It would characterize average deviation of GB from bisector direction. To include the effect of separation between defects we introduced relative length of defect $d = D/D_0$, where D is the distance between centers of two consecutive defects in the GB and $D_0$ is a distance between the centers of defects in a perfect 5-7 pair where pentagon and heptagon are adjacent. The data for length normalized absolute value of directional angle $<|γ|*d>$ are shown on **Fig.4b**.

We also calculated the Herman's orientation function $S_d = (3\cos^2 γ - 1)/2$ (**Fig. 4c**), commonly used to characterize alignment of structures [30]. Value of 1 corresponds to defect pairs parallel to bisector of β, 0 to disordered structure, and -0.5 to defect couples perpendicular to the bisector. This measure allows us to determine the region β = 38.2 ÷ 81.8 degrees for which straight and scattered straight structures of GB are characteristic.

To characterize the continuity of GB we analyzed the average inverted relative length $<1/d>$ of defects in structure (**Fig.4d**). Value of $<1/d> = 1$ corresponds to a continuous structure in which each previous defect is in contact with the next one. As expected, continuous GBs correspond to the regions around α ≈ 30° (β ≈ 30°, 60°).

In addition we plot average energy per length of GBs (**Fig.4e**). Structures were optimized both in planar-confined (2D) and free 3D conditions. In both calculations constraints were imposed to preserve misorientation angle unchanged.

Further analysis of the data allow us to distinguish 6 regions of edge misorientation angle β for which GB are differ in their characteristics.

*I: β = 0 ÷ 21.8° (α = 0° ÷ 21.8°).* Structures in this region are alternating (first rotation and center peak orientations), non-continuous and, interestingly, straightly periodic. Periodicity of GBs in this region is explained by confined conditions of their growth. As soon as one defect pair is created during growth it automatically sets conditions for creation of the next pair and therefore no deviation is occurring.

*II: β = 21.8° ÷ 38.2° (α = 21.8° ÷ 30.0°).* Structures in this region are alternating (center peak and first rotation orientations of defect pairs) and continuous. Energies of GBs correspond to one of the two minima (α ≈ 30) which are usually associated with high density of defects and orientation of defect pairs that cause cancelation of stress in GB.

*III: β = 38.2° ÷ 60° (α = 21.8° ÷ 0.0°).* GBs have straight or scattered straight (center peak and shift defect pair orientation), non-continuous structures. The only truly straight structure in this region corresponds to β = 38.2 (α = 21.8) and is periodic. That phenomenon is explained by strong interaction between defect pairs during formation of GB. Energy of GBs decrease to virtually zero value at β = 60 (α = 0) as no tilt GB is formed.

*IV: β = 60° ÷ 81.8° (α = 0.0° ÷ 21.8°).* This region is nearly symmetric to region III. Structures are straight at β = 81.8 (α =21.8) and scattered straight on the rest of the region (center peak and shift orientations of defect pairs). GBs are non-continuous.

*V: β = 81.8° ÷ 98.2° (α = 21.8° ÷ 0.0°).* Structures are alternating and curved (center peak, first rotation and second rotation orientation of defect pairs) and continuous. That is the only region with curved GB (second rotation of defect pairs — see Fig. 3d,i). Similar shapes of GB were observed experimentally by Huang *et al.* [23] and Kim *at al.* [24].

*VI: β = 98.2° ÷ 120° (α = 21.8° ÷ 30.0°).* In general this region is similar to region I, but due absence of confined growth conditions additionally shift orientation of defect pairs is present in GBs.

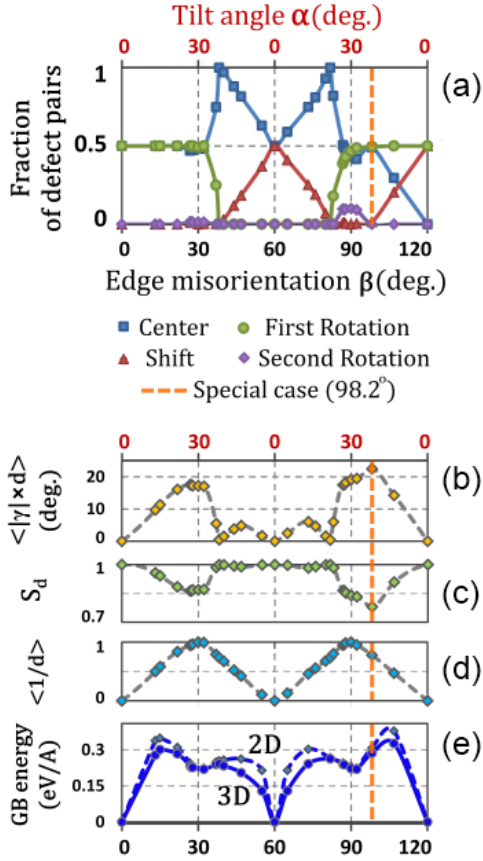

**Figure 4. Quantification of GB wiggliness.**
(a) Structural composition of GB expressed as proportion between four basic types of defect pair orientation. Values for special case of 98.2° (21.8°) are explained later. (b) The average length normalized absolute value of directional angle <|γ|*d>. Regions with <|y|*d> = 0 correspond to structures with no tilt GB. (c) Herman's orientation function $S_d = (3*cos2|γ| - 1) / 2$. Region with $S_d$ = 1 corresponds to structures with straight GB. (d) Average inverted relative length of defects < 1/d > as measure of continuity of GB. Regions with < 1/d > = 1 correspond to continuous GB. (e) Energy of GB (eV/A) optimized in 2D and 3D.

Therefore GBs are alternating and scattered alternating (center peak, shift and first rotation orientations of defect pairs) and discontinuous. GB structure for *β = 98.2° (α = 21.8°)* stands out from among the others. Unlike the rest of the GBs it corresponds to the same misorientation angle as two very periodic and structurally rigid GBs (alternating for β = 21.8° and straight for β =38.2°). The resulting structure of GB (Fig. 5a) is composed of regions of alternating GB following bisector direction and straight GB following first rotation direction. Structural content of the resulting structure as well as both separated regions is shown on **Fig. 5b**. It's important to mention that value for β = 98.2° on Fig. 4a corresponds to the only alternating structure of GB that follows bisector direction; we plot value for alternating GB as that is the structure that is consistent with the rest of the region VI, while regions of straight GB following first rotation direction are not present in any other GB of this region.

In general we can conclude that structures with identical value of misorientation angle α can correspond to significantly different structures of GB if those were formed from different initial conditions (edge misorientation angle β). And therefore edge misorientation angle β is an essential parameter in case one needs to determine initial conditions of growth from final structure of GB or determine required initial grains orientation to form specific type of GB.

## "Translational" grain boundaries

For β = 0°, 60° and 120° (α = 0°) no tilt grain boundaries are formed as grains virtually belongs to the same lattice. However if grains are shifted from a perfect position then translational GB would be formed. This could occur in cases such as Ni(111) substrate or any other surface that enforces a preferential orientation on graphene islands due to its symmetry [31]. Translation in this case is possible in two directions: along initial edges (Δy) and along growth direction (perpendicular to edges – Δx). We simulate formation of GB for the full range of shift in both directions as well as for the full range of combinations of shift directions; the scheme of the experiment is shown in Figure 6 (a). Several of the resulting structures are shown on **Fig. 6 (b-e)**. Using all observed structures we composed the energy surface of structures corresponding to the full range of Δx and Δy.

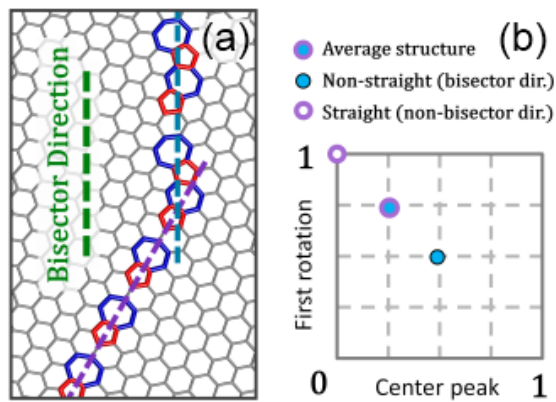

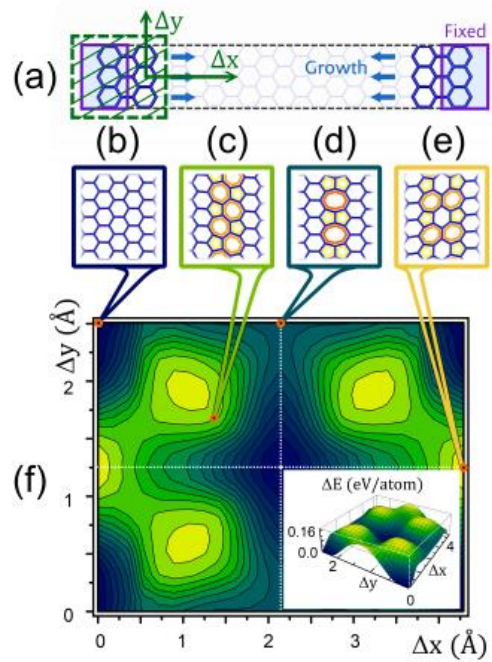

**Figure 5. Special case: grain boundary for β = 98.2° (α = 21.8°). (a)** Structure of grain boundary for β = 98.2° is essentially a mixture of two types of GB. First type is an alternating structure (as found for β = 21.8°) parallel to bisector direction. Second type is a straight structure (as found for β = 38.2° and 81.8°) which follows direction at 30° degrees from bisector direction. (b) Structural content of average structure as well as both separate types is shown for comparison.

**Figure 6. Translational grain boundaries. (a)** Scheme of the computational experiment for translational GB (β = 0°): GB is formed as a result of one grain being shifted from the position on the perfect lattice of the second grain. Shift direction can be represented as combination of two principal directions: perpendicular to the initial edge (Δx) and parallel to the initial edge (Δy). **(b-e)** Sample structures corresponding to some shift values. **(f)** Energy of structures corresponding to α = 0° with different shift.

As expected, the lowest energy value corresponds to zero translation in both directions - perfect graphene structure (**Fig. 6b**). A perfect structure without a boundary also corresponds to a half shift in both directions $\Delta x = \Delta x_{max}/2$, $\Delta y = \Delta y_{max}/2$ (center point of the plot). The structure with consequent pairs of defects oriented perpendicular to the GB (**Fig.6c**) was observed in experimental work by Biachini *et al.* [32]. The structure consisting of octagons separated by pairs of pentagons (**Fig. 6d**) was reported by Lahiri *et al.* [33]. We found the same structure to correspond to a saddle point of energy surface at $\Delta x = \Delta x_{max}/2$ and $\Delta y = 0$. Both structures (c) and (d) correspond to the same amount of in-plane shift (one bond length) but at different angle compared to the edge of the grain. Thus, they correspond to the same relative position of original grains and may potentially convert one into the other upon turning. The structure with the highest energy corresponding to the half shift in y direction ($\Delta x = 0$, $\Delta y = \Delta y_{max}/2$) is shown in **Fig. 6e**. Once again we see that the geometrical conditions (shifts in this case) and the corresponding local energy preferences during growth override the global minimization of energy.

## Strategies of morphology control

Our findings provide a theoretical base to choice of grains orientation prior to growth in order to create desired structure (e.g., to minimize the scattering of electrons or to induce local metallicity or valley filtering effects). As a simple application, substrates with a sixfold or twofold symmetry—(111) or fcc (110)–that enforce preferred orientations with β = 0/60/120° will tend to form regular untilted structures, while e.g. fourfold-symmetric surfaces such as fcc (100) will add β = 30/90° to the mix [31], producing irregular boundaries.

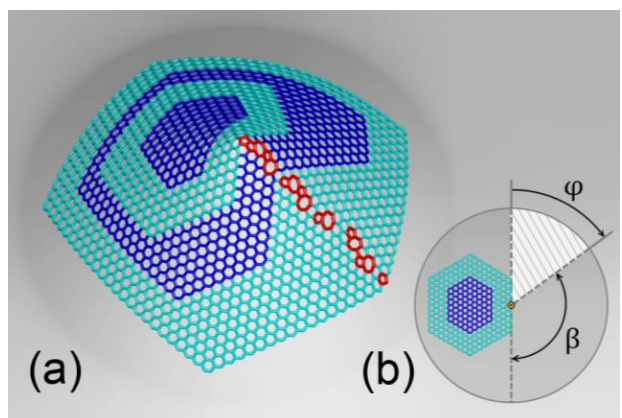

**Figure 7. Graphene growth with formation of GB on convex substrate. (a)** Temporal evolution follows from the initial island (usual zigzag-edge hexagon [27], dark blue) to light-blue still hexagon, and upon passing over the prism or cone apex it transforms into a (dark-blue) heptagon, which eventually converges into all-zigzag edge pentagonal island, while a grain boundary forms (red). The edge misorientation angle of the heptagon is determined by the cone disclination angle φ (α = φ modulo 30°) **(b)**.

Our recent study shows how the semi-infinite GBs induce off-plane conical warping [34] and suggests another means of tuning β over a continuous spectrum of values by using 3D-patterned substrates, e.g., with conical topography as illustrated in **Fig. 7a**. The angle of the cone θ is related to the disclination angle of the cone φ as: θ = 2 arcsin (1 − φ/2π) [35]. For φ = n·π/3 we recover the five canonical cones with no GB. Otherwise a GB is formed, running from the apex of the cone to its base. It is easy to show that the edge misorientation angle is then determined as β = π − φ = π − 2π(1 − sin $^{θ}/_{2}$). Interesting, as seen in **Fig. 7b**, is the non-monotonous evolution of the growing island shape: from a small island shaped as common zigzag terminated *hexagon*, to the *heptagon* (after it "breaks" to adapt to the sharp apex), and then back to *pentagon*, but now containing a GB of a prescribed tilt, as 6 → 7 → 5 sequence.

## Summary

We present analysis of the kinetically formed grain boundaries in graphene created using a Monte Carlo method. Simulation of grain boundary formation allowed identification of a hidden degree of freedom - edge misorientation β. GB growth simulation was performed for the full range of β (and therefore α). We determine four types of GB shape and defect pairs orientations corresponding to those GB shapes. Analysis of observed results showed that straight (predicted theoretically) and scattered straight structures are formed for β = 32.8° ÷ 81.8° (region is nearly symmetric around β = 60°). For edge misorientation within 0° ÷ 21.8° and 98.2° ÷ 120° regions GBs have non-symmetric structure even though they end up having a higher energy than straight structures. For a special value β = 98.2° structure was found to be formed by regions of non-symmetric GB and regions of symmetric GB combined. Continuous alternating structures are observed around β ≈ 30°. Curved structures commonly observed in experiments were found to be formed only around β = 90°. For zero misorientation (β = 0°, 60°, 120°) no tilt GBs are formed but translational GB are possible due to shift of grains from positions on the same lattice. Presented work provides precise analysis of dependency of GB structure from initial conditions of growth (grains orientation). We propose strategies to control the morphology of GB in graphene, for example by a suitable choice of substrate symmetry or by using 3D-patterned surfaces. Conversely, our findings can be used to analyze growth conditions from structure of resulting GB.

## Supporting materials

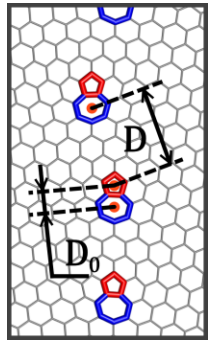

**Figure S1. Definition of relative length of defect pair – d.** Relative length of defect pair is defined as relation between length of defect pair under consideration D to length of defect pair with pentagon and heptagon adjacent to each other $D_0$ (d= $D/D_0$).

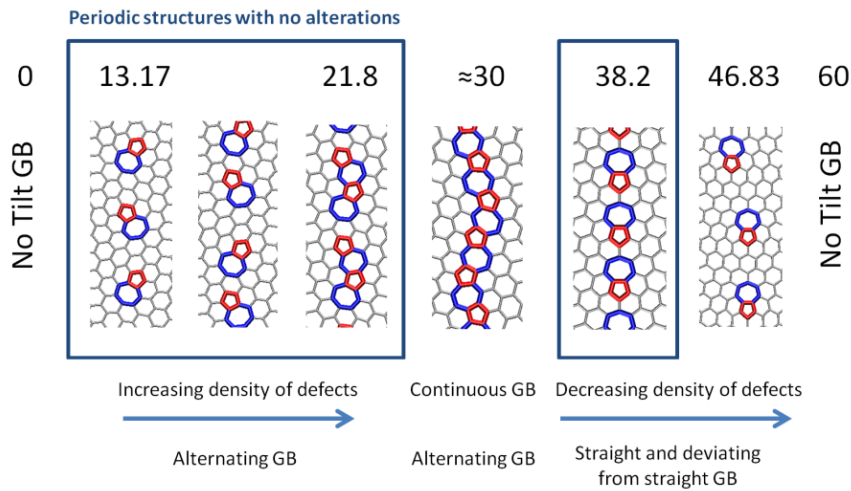

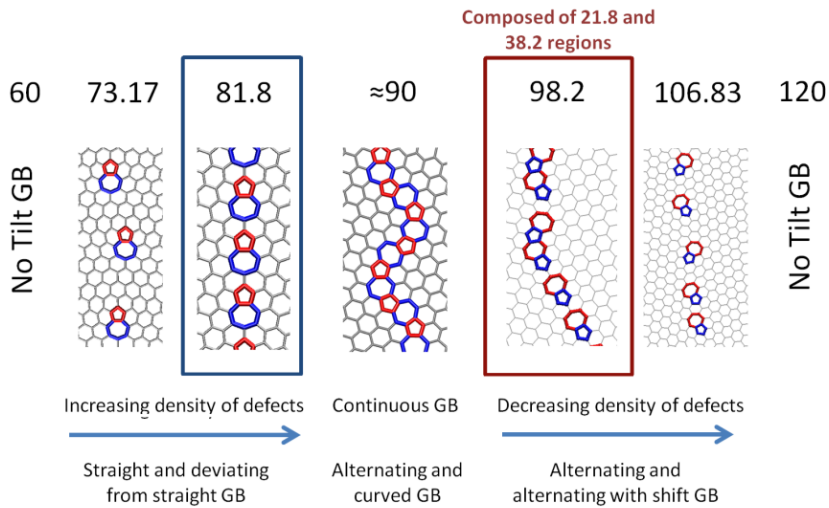

**Figure S2. Typical structures of GB for the complete range of edge misorientation angle β.**